\documentclass[twocolumn]{aastex6}

 \usepackage{appendix}
\usepackage{graphicx}
\usepackage[percent]{overpic}
\usepackage{txfonts}
\usepackage{xcolor}
\usepackage{pict2e}
\usepackage{pbox}
\usepackage{booktabs}
\usepackage{multirow}
\usepackage{color, colortbl}
\definecolor{Gray}{gray}{0.9}
\usepackage{placeins}
\usepackage{hyperref}
\begin{document}

\title{Differences in periodic magnetic helicity injection behaviour between flaring and non-flaring Active Regions: Case Study}

\author{M. B. Kors\'os\altaffilmark{1,2}, P. Romano\altaffilmark{3},  H. Morgan\altaffilmark{1}, Y. Ye \altaffilmark{4}, R. Erd{\'e}lyi\altaffilmark{5,2}, F. Zuccarello\altaffilmark{6} }

\altaffiltext{1}{Department of Physics, Aberystwyth University, Ceredigion, Cymru, SY23 3BZ, UK}
\altaffiltext{2}{Department of Astronomy, E\"otv\"os Lor\'and University, P\'azm\'any P\'eter s\'et\'any 1/A, H-1112 Budapest, Hungary}\
\altaffiltext{3}{INAF Osservatorio Astrofisico di Catania, Via S. Sofia 78, I 95123 Catania, Italy}
\altaffiltext{4}{Space Science Institute, Macau University of Science and Technology, Macao, People's Republic of China}
\altaffiltext{5}{Solar Physics \& Space Plasma Research Center (SP2RC), School of Mathematics and Statistics, University of Sheffield, Hounsfield Road, S3 7RH, UK}
\altaffiltext{6}{Dipartimento di Fisica e Astronomia ``Ettore Majorana", Universit\`a di Catania, Via S. Sofia 78, I 95123 Catania, Italy}

\begin{abstract}
{The evolution of magnetic helicity has a close relationship with solar eruptions and is of interest as a predictive diagnostic. In this case study, we analyse the evolution of the normalised emergence, shearing and total magnetic helicity components in the case of three flaring and three non-flaring  active regions (ARs) using SHARPs (Spaceweather Helioseismic Magnetic Imager Active Region Patches) vector magnetic field data. The evolution of the three magnetic helicity components is analysed with wavelet transforms, revealing significant common periodicities of the normalised emergence, shearing and total helicity fluxes before flares in the flaring ARs. The three non-flaring ARs do not show such common periodic behaviour. This case study suggests that the presence of significant periodicities in the power spectrum of magnetic helicity components could serve as a valuable precursor for flares.}
\end{abstract}


\section{Introduction}

Space weather refers to the short-term interaction of different manifestations of solar activity with geospace that occurs through a complex series of dynamic events. These interactions can result in hazardous conditions for the functioning of many vital socioeconomic infrastructures, both terrestrial (e.g., long-distance oil/gas pipelines, electric power grids, aviation-control, HF radio communication) and space-based (e.g., communication satellites, global positioning systems, ISS), leading to a reduced or total lost capacity \citep[][and references therein]{Eastwood2017}. Advancements of solar eruption forecasting capabilities through the identification of observable precursors at the Sun is crucial \citep[see e.g.][and references therein]{Barnes2016,Leka2019}. This forecasting is challenging, in particular understanding the physical processes that underpin solar eruptions \citep[see e.g.][]{Florios2018, Campi2019, Korsos2019, Yimin2019}. 

Flares and coronal mass ejections (CMEs) originate mostly from magnetically complex, highly twisted and sheared elements of a $\delta$-type active region (AR) \citep[e.g.][and references therein]{Toriumi2019, Georgoulis2019}. The evolution of the magnetic helicity \citep{Elsasser1956} is likely to be a key physical process that precedes flare and/or CME events, and measurements of helicity derived from photospheric magnetic field data can provide insight into the underlying mechanism(s) of these events. Many observational studies have found a relationship between the temporal evolution of helicity flux and flares/CMEs. 

\cite{Moon2002a,Moon2002b} found that a significant amount of helicity is injected before large flare events. \cite{Smyrli2010} investigated the helicity flux in a case study of 10 ARs and reported a sudden change in the helicity flux was present during six flares. \cite{Park2008, Park2012} discovered that the helicity flux slowly increases and then remains constant just before flares. Park {\it et al.} also reported that the injected helicity flux changed its sign before some very impulsive flare and CMEs. \cite{Tziotziou2013} studied the dynamic evolution of AR 11158 before flares and CMEs, and found that eruption-related decreases, and subsequent free-energy and helicity budgets, were consistent with the observed eruption magnitude.
Other works addressed also that the helicity flux reversed sign around at the start of a flare \citep{Vemareddy2012,Wang2014,Gao2018}, caused by the interaction between the associated magnetic flux tubes with opposite signs of helicity \citep{ Linton2001, Kusano2004, Liu2007, Chandra2010, Romano2011a, Romano2011b}. It is suggested that a CME can also remove helicity from its source, leading to a lower total AR potential magnetic energy \citep{Demoulin2002, Smyrli2010}. On the other hand, based on numerical data, \cite{Pariat2017} claimed that magnetic energies and the total relative helicity are not effective diagnostics for flare prediction, but the decomposition of the relative magnetic helicity introduced by \cite{Berger2003}, in the current-carrying component and its counterpart, may be useful. Based on solar magnetic field observations, \cite{Thalmann2019} gave similar conclusions to \cite{Pariat2017}. They reported that the ratio of current-carrying to total helicity is capable of indicating an eruptive AR, but not the magnitude of an upcoming eruption.

\begin{figure*}[t!]
\centering
\includegraphics[width=1\textwidth]{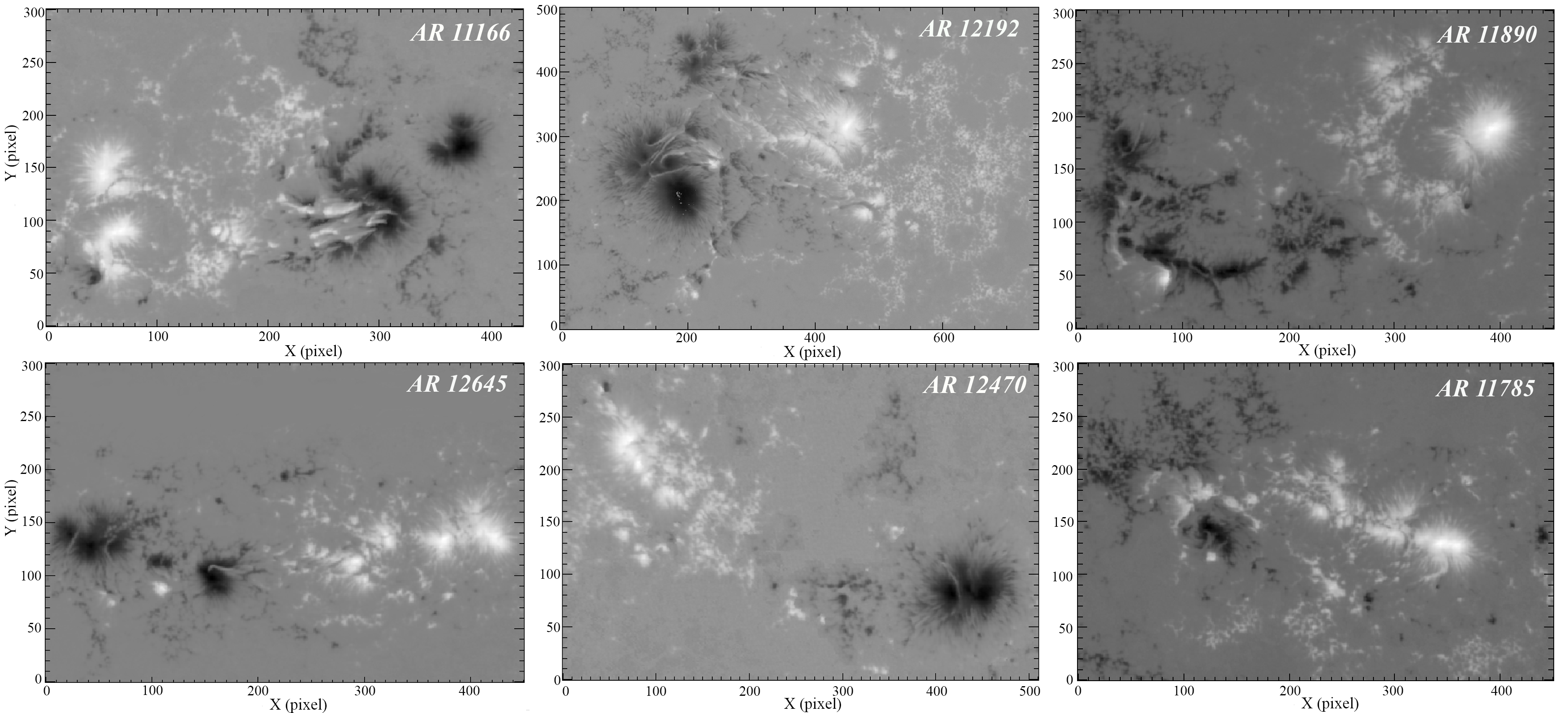}
\put(-495,220){\textcolor{white}{(a)}}
\put(-325,220){\textcolor{white}{(b)}}
\put(-155,220){\textcolor{white}{(c)}}
\put(-495,105){\textcolor{white}{(d)}}
\put(-328,105){\textcolor{white}{(e)}}
\put(-155,105){\textcolor{white}{(f)}}

\caption{\label{ARs}  
Six panels showing the radial component of magnetic field in (a) AR 11166 on 2011.03.10 20:24:00, (b) AR 12192 on 2014.10.24 00:00:00, (c) AR 11890 on 2013.11.08 00:00:00, (d) AR 12645 on 2017.04.02 00:00:00, (e) AR 12470 on 2015.12.17 00:00:00, and (f) AR 11785 on 2013.07.09 00:00:00. The {\it x} and {\it y} axes are expressed in pixels, where one pixel is related to the 0.5 arcsec resolution of SDO/HMI.}
\end{figure*}
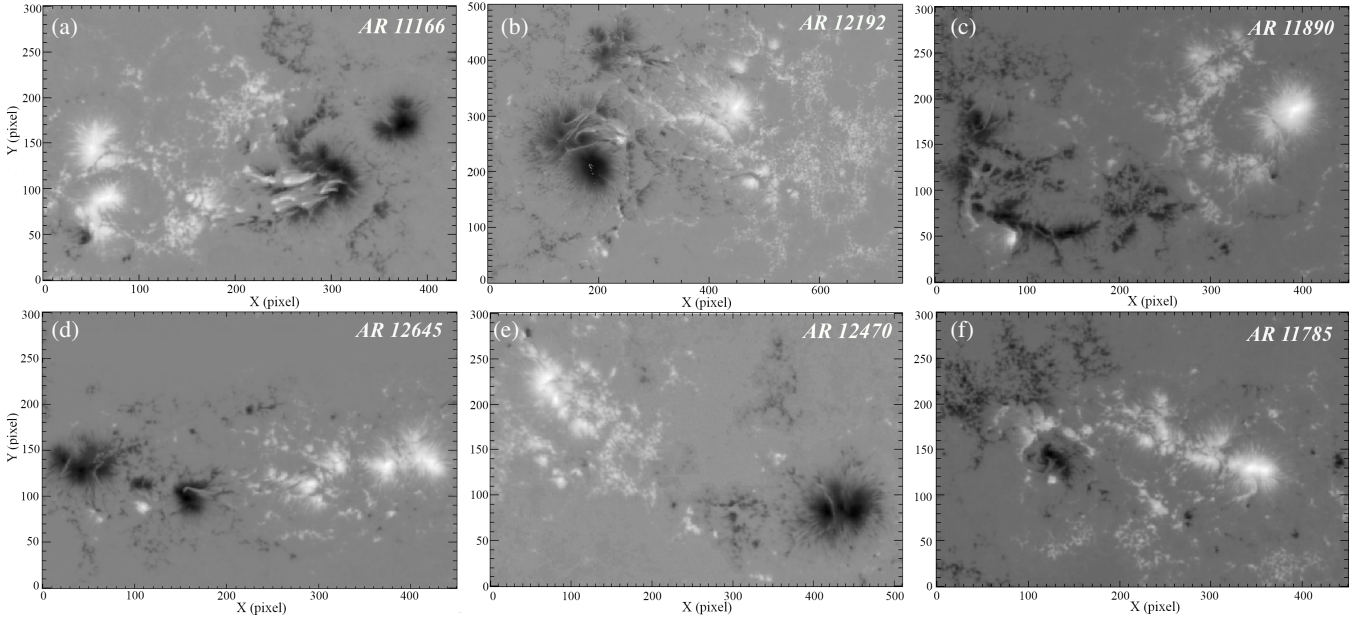

To the best of our knowledge, no previous studies have found a relationship between the oscillatory behaviour of the evolution of magnetic helicity (or its various components) and flaring activities that may be related to such oscillations. A recent theoretical study by \cite{Prior2020} reported that the multi-resolution wavelet decomposition is useful to analyse the spatial scales of helicity in magnetic fields in a manner which is consistently additive.
To investigate distinctive behaviour patterns of helicity flux in flaring and non-flaring ARs as a case study to demonstrate the concept, we focus here on the evolution of the magnetic helicity injection rate through wavelet analyses during the observable disk passage period of six ARs. 

The work is structured as follows: Section~\ref{data} describes the adopted tools for the helicity flux calculations and the selection criteria of these six investigated ARs. In Section~\ref{analyses}, we describe the analysis, and present the results. Key findings and conclusions, along with a suggestion of future work, are given in Section~\ref{conclusion}. 

\section{Magnetic helicity flux calculation}  \label{data}

The magnetic helicity is a proxy for the 3D complexity of a magnetic field in a volume, thus, it is often interpreted as a generalisation of more local properties such as magnetic twist and shear. \cite{Taylor1974} and \cite{Woltjer1958} introduced the concept of magnetic helicity as a well-conserved quantity, even in non-ideal magnetohydrodynamics. \cite{Berger1984b} proved that helicity is conserved in conductive plasma, meaning that helicity variations with respect to time are essentially restricted to helicity flow through a surface $S$. \cite{Berger1984b} showed that magnetic helicity dissipates very slowly during the course of magnetic reconnection. 

To monitor the helicity flux (i.e., the helicity injection rate) through the photosphere over an AR, we use the following equation:  
\begin{equation} \label{Origin}
\left. \frac{dH}{dt}\right|_{S}=2\int_{S}\left( {\bf A}_{p}\cdot {\bf B}_{h}\right){\bf v}_{\bot z}dS-2\int_{S}\left( {\bf A}_{p}\cdot {\bf v}_{\bot h}\right) {\bf B}_{z}dS,
\end{equation}
introduced by \cite{Berger1984}. ${\bf A}_{p}$ is the vector potential of the potential magnetic field ${\bf B_{p}}$. ${\bf B}_{h}$ and ${\bf B}_{z}$ denote the tangential and normal components of the magnetic field vector with respect to the surface S, and ${\bf v}_{\bot h}$ and ${\bf v}_{\bot z}$ are the tangential and normal components of velocity. The first term on the right side arises from twisted magnetic flux tubes emerging from the solar interior into the corona (emergence term hereafter), while the second term is generated by the shearing and braiding of the field lines by tangential motions on the solar surface (shearing term hereafter). 
 ${\bf A}_{p}$ is determined by the photospheric magnetic field and the Coulomb gauge \citep{Berger1997,Berger2000}. 

  \begin{figure*}[t!]
\includegraphics[width=1.01\textwidth]{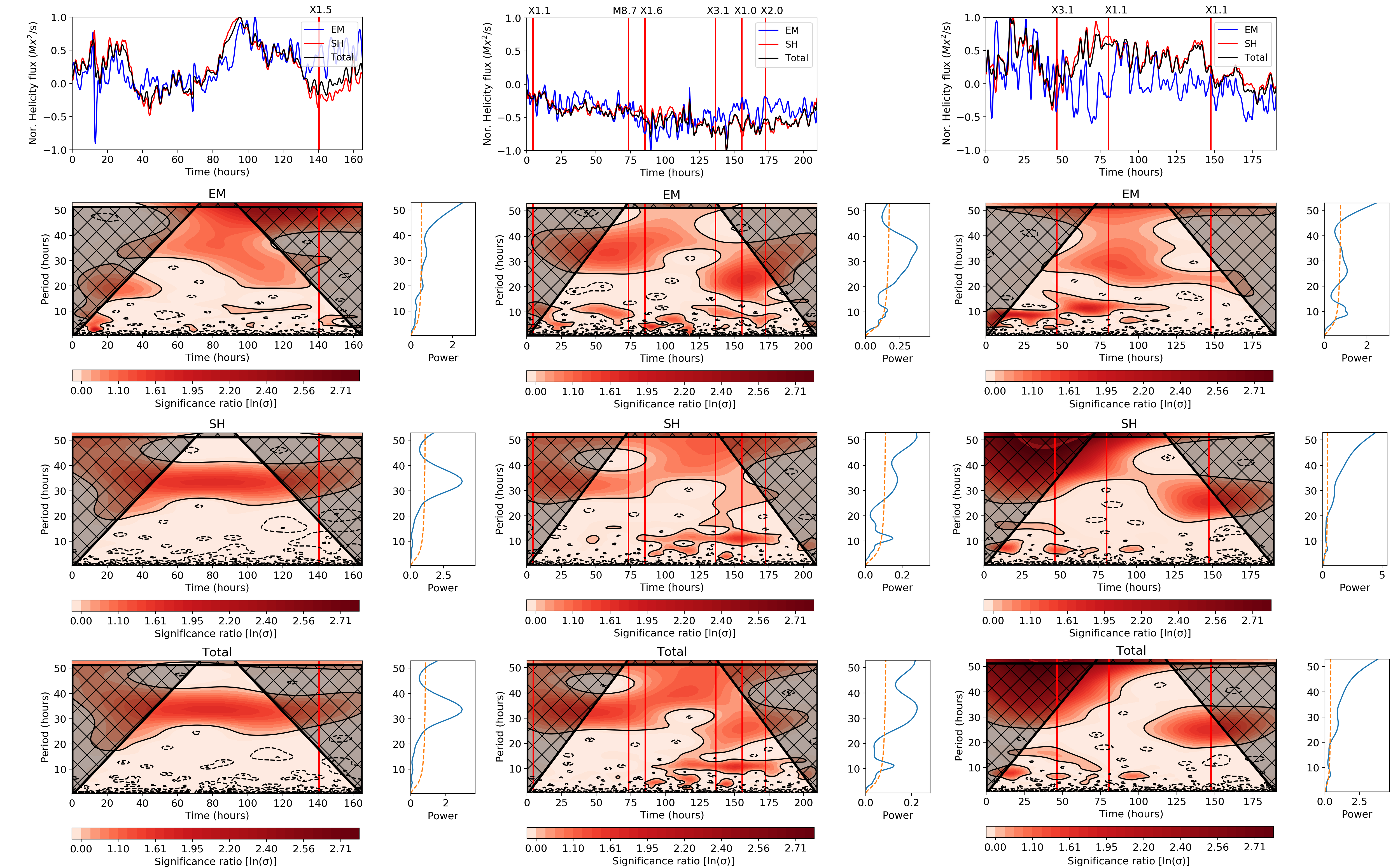}
\put(-510,323){(a)}
\put(-455,323){ {\bf AR 11166}}
\put(-340,323){(b)}
\put(-295,323){ {\bf AR 12192}}
\put(-170,323){(c)}
\put(-125,323){ {\bf AR 11890}}
\caption{\label{Flare}  
Time analysis of three flaring ARs, namely (a) AR11166 (b) AR 12192, and (c) AR 11890. The top panels show time series of the normalised emergence (EM, blue), shearing (SH, red), and total (Total, black) helicity fluxes. The red vertical lines mark the onset time of flares. The 2$^{nd}$-4$^{th}$ rows show the wavelet power spectrum (WPS) of the normalised emergence, shearing, and total helicity fluxes, respectively. Rather than plotting power directly, the colour bars visualize the logarithm of the ratio of the power to the expected power for Gaussian-distributed white noise and significance compared to noise. The $x$-axis of each WPS is the observation time, and the $y$-axis is the period, both in hours. On the WPS plots, the black lines bound the cone-of-influence, i.e. the domain where edge effects become important. The plots to the right of each WPS are the corresponding global wavelet spectra (power averaged over time, similar to a Fourier power spectrum). The orange dashed lines mark the one $\sigma$ confidence in global wavelet spectrum analyses.}
\end{figure*}

For magnetic helicity calculation, a reliable and continuous photospheric vector magnetograms are required in order to determine the associated photospheric velocity fields. Therefore, we use $hmi.sharp\_cea\_720s$ vector magnetic field measurements of the Spaceweather Helioseismic Magnetic Imager Active Region Patches (SHARPs\footnote{http://jsoc.stanford.edu/doc/data/hmi/sharp/sharp.htm}), with a 12-min cadence \citep{Bobra2014}. The photospheric plasma velocity is calculated by applying the Differential Affine Velocity Estimator for Vector Magnetograms (DAVE4VM\footnote{https://ccmc.gsfc.nasa.gov/lwsrepository/DAVE4VM\_description.php}) algorithm \citep{Schuck2008}. The window size used in the calculations is 19 pixels, which was determined by examining the non-parametric Spearman rank order correlation coefficients, Pearson correlation coefficients and slopes between $\Delta_{h} \cdot ({\bf v}_z{\bf B}_h-{\bf v}_h{\bf B}_z)$ and $\delta {\bf B}_z/\delta t$ \citep{Schuck2008}. The vector potential ${\bf A}_{p}$ is derived using MUDPACK \citep[for details see e.g.][]{Adams1993}, a multigrid software for solving elliptic partial differential equations.

In this case study, we analyse the magnetic helicity flux evolution of six ARs, namely NOAA  ARs 11166, 11785, 11890, 12192, 12470 and 12645 (see Fig.~\ref{ARs}), which were selected to satisfy the following criteria: 

\begin{itemize}
\item The selected ARs respect the Hale-Nicholson law \citep{Hale1925} of solar cycle 24. Some works claim that AR that violate the Hale–Nicholson law are more flare/CME productive \citep[][and references therein]{Elmhamdi2014}.
\item The ARs have a prevalent bipolar configuration.
\item The ARs have $\delta$-spot(s). 
\item  The ARs have two distinct behaviours in terms of flare activity. ARs 11166, 11890 and 12192 were host of intense M and X-class flares, and are grouped as ``flaring''. ARs 11785, 12470 and 12645 only produced B- and C-class flares and are grouped as ``non-flaring''.
\item The AR is not a cradle of significant/fast CME eruptions that have linear velocities larger than $\sim$1000 km/s. In this regard, ARs 11166 and 11890 produced slow CMEs only, with 400-700 km/s linear speeds. AR 12192 was rich in terms of flaring but not in terms of CMEs.
\item In each of the flaring/non-flaring groups, one-one AR is characterised to be dominantly either in formation (ARs 11166 and 12645), or fully developed (ARs 12192 and 12470), or in a decaying (ARs 11890 and 11785) evolutionary phase during the investigated period.
\end{itemize}

Table~\ref{ARsmagnetogramtable} summarises the time interval of observations of the six ARs. Each time interval is limited to a duration when the corresponding AR is between $-60^{\circ}$ and $+60^{\circ}$ with respect to the central meridian to avoid extreme magnetic field projection effects \citep{Bobra2014}. Table~\ref{ARsmagnetogramtable} includes also the onset time and the associated GOES class of the flares occurred in each AR, based on the GOES solar flare catalog\footnote{https://hesperia.gsfc.nasa.gov/}. Furthermore, Table~\ref{ARsmagnetogramtable} gives information on the dominant evolutionary phases of the ARs, and on how long their $\delta$-spots were observed. 

\section{Data Analysis} \label{analyses}

\begin{figure*}[t!]
\includegraphics[width=1.01\textwidth]{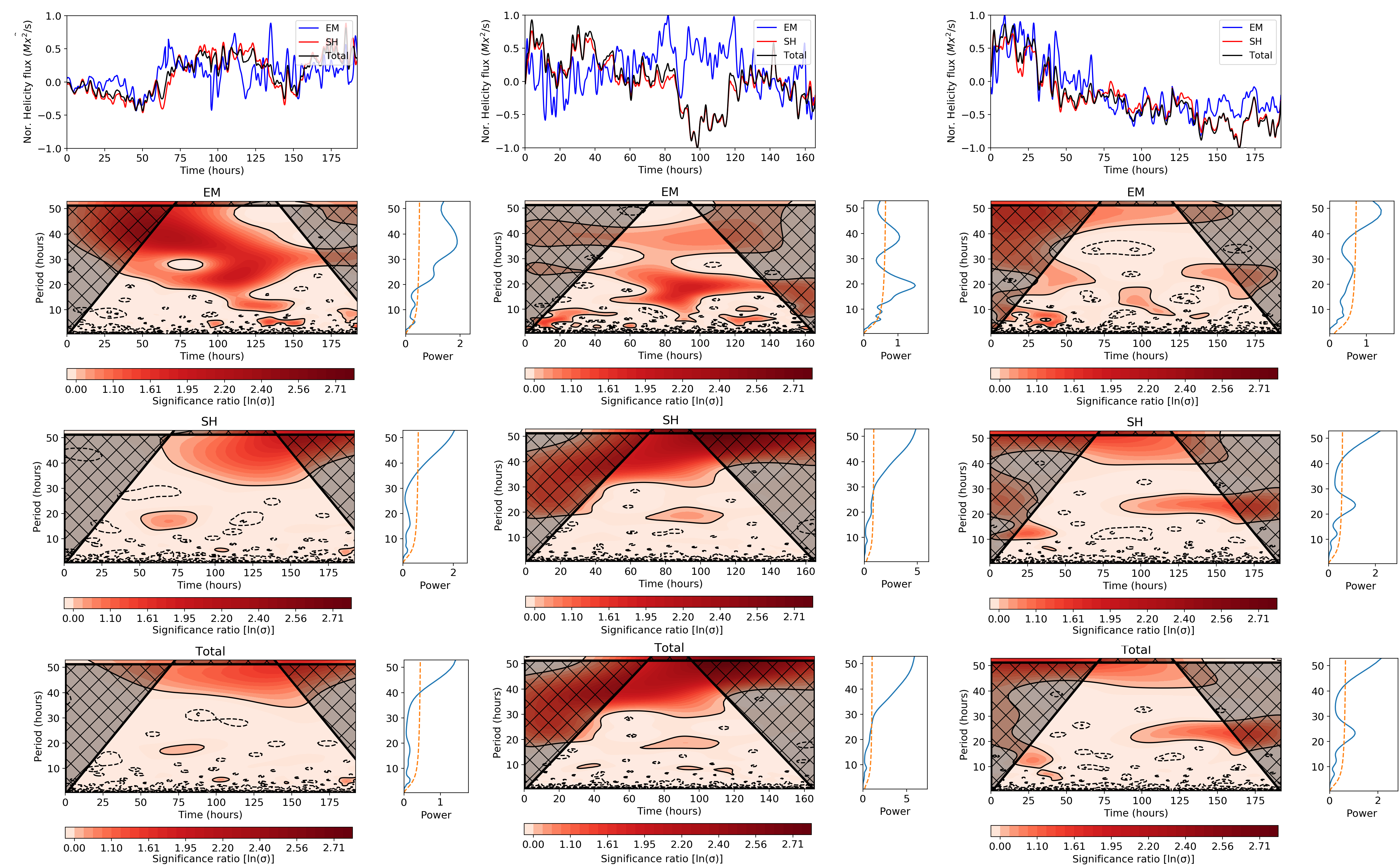}
\put(-510,320){(a)}
\put(-455,320){ {\bf AR 12645}}
\put(-340,320){(b)}
\put(-295,320){ {\bf AR 12470}}
\put(-170,320){(c)}
\put(-125,320){ {\bf AR 11785}}
\caption{\label{No-Flare}  
Same as Fig.~\ref{Flare}, but for the three non-flaring ARs, namely (a) AR 12645, (b) AR 12470, and (c) AR 11785.}
\end{figure*}

The emergence and shearing components of the magnetic helicity of the six ARs are calculated using Equation (\ref{Origin}). The total magnetic helicity flux for a given AR was generated by summing the emergence and shearing components. The three helicity injection rates for each AR were obtained by integrating the helicity flux over the entire area of the AR. The three helicity flux components are further normalised by their respective largest absolute value in order to facilitate comparison on similar scales. The normalised emergence, shearing and total helicity fluxes are shown in the top panels of Figures ~\ref{Flare}--\ref{No-Flare}. In the upper panels of Figs.~\ref{Flare}--\ref{No-Flare}, the blue line is the emergence term, the red line the magnetic helicity flux associated with shearing motions at the photosphere, and the black line is the total magnetic flux. 

By inspecting the top panels of Figures~\ref{Flare}--\ref{No-Flare}a-c, we can see that the normalised shearing and total helicity flux components show similar evolution trends in each of the six AR cases, which suggests that the shearing motion plays a more important role in the evolution of total helicity.
The emergence helicity flux develops differently when compared to the two other components, see e.g.\ in case of AR 11890 or 12470. Furthermore, we can also identify various quasi-periodic patterns in the evolution of the three helicity components, for both flaring and non-flaring AR cases. However, to reveal a possible diacritical periodic signal(s) between the two groups, we construct wavelet power spectra (WPS) using a software developed by \cite{Torrence1998}, employing the default Morlet wavelet profile. The associated global power spectrum (GPS) is also calculated as the WPS averaged over time for each case. This is similar to a Fourier power spectrum. The significance level of the WPS, at one $\sigma$ confidence level, is estimated using a white noise model and the standard deviation of the input signal. This significance is a function of the periodicity. Therefore, the ratio of the WPS to the significance level is useful to identify significant periodicities - we call this value the significance ratio. In Figures ~\ref{Flare}--\ref{No-Flare}a-c, the natural logarithm of this ratio is displayed, therefore values of 0 or higher (or within the black contours) are significant.

In Figures~\ref{Flare}--\ref{No-Flare}a-c, the WPS and GPS of the emergence (EM), shearing (SH) and total (Total) helicity fluxes are shown, after application of a high-pass filter. The original data series is smoothed with a time scale of two-third of the full length of the time series, and the resulting smoothed series is subtracted from the original data series, thus damping power at long periods. This is an important step because slow changes that are not of immediate interest for this study may affect the calculation of significance levels at shorter periods \citep{McAteer2002}.
In this study, a significant period is identified as (i) a significance ratio larger than 1 (i.e. that is 0 on the {\it ln} scale) measured in $\sigma$, and (ii) the peak in the GPS is above the confidence level (shown as the dashed orange line in the GPS plots).

Based on Figures~\ref{Flare}a-c, in the case of three flaring ARs, there are common peak(s) of the EM, SH, and total helicity flux components in the WPS and GPS preceding the flare occurrences. In particular:
\begin{itemize}
\item {\it AR 11166}: In Fig.~\ref{Flare}a, a powerful the 34-hr periodicity is present in the GPS of the EM, SH, and total helicity fluxes about 5 days prior to the X1.5 flare occurrence. The WPS shows that 34-hr periodicity persists for a lifetime of  $\sim$3 cycles in the EM time series and declines after the flare. However, this periodicity continues to play an important role for 5 cycles in the SH and total helicity fluxes. The EM shows a 20-hr short-lived periodicity (2 cycles) near the start of the time series. This feature is a result of the abrupt large negative value in EM that is not present in the data of SH and in the total flux components. It is interesting to note that the 20-hr periodicity coincides with the large variations (and even a change in sign) in the EM, and could be related to the findings of \cite{Smyrli2010} and/or \cite{Park2008, Park2012} mentioned in the Introduction.

\item {\it AR 12192}: There are two common peaks in periodicities in the three helicity flux components, see Fig.~\ref{Flare}b. First, a $\sim$35-hr strong periodicity is observable before the M8.6 and X1.6 flares in the EM, SH, and total flux time series, as shown in both the WPS and GPS. After the M8.6 and X1.6 flares, a $\sim$10-hr periodicity also becomes dominant next to the $\sim$35-hr prior to the remaining three X-class flares. This $\sim$10-hr common periodicity is sporadically present in the EM and SH time series. However, this peak appears only after the X1.6 flare in the evolution of the total flux data. The lifetime of the $\sim$35-hr periodicity is longer than 5 cycles in the case of the three helicity flux components. Nevertheless, the $\sim$10-hr period is observed through 20 cycles in the case of the EM, while this periodicity is continued during 13/11 cycles in the SH/total data.

\item {\it AR 11890}: Similar to AR 12192, we identify common peaks at two distinct periodicities in the three helicity components prior to the X-class flares in Fig.~\ref{Flare}c. The first peak period is $\sim$8-hr, which appears before the X3.3 and the first X1.1 flare, respectively. This $\sim$8-hr period decays after 10 cycles, just after the first X1.1 event. In the case of the second X1.1 flare, the $\sim$28-hr periodicity peak becomes a common feature only from $\sim$115 hrs in the WPS of the EM, SH and total flux components. The $\sim$28-hr peak appears earlier and is observable throughout 5 cycles in the EM, compared to a lifetime of only 3 cycles in the SH and total flux components.

 \end{itemize}

%Next, let us now investigate the wavelet and global power spectra of the non-flaring ARs. 
Common significant periodicities are absent in the helicity flux components of the three non-flaring AR. Only the EM flux time series of ARs 12470 and 12645 show some peaks in the WPS and GPS of Fig.~\ref{No-Flare}.
 In the case of AR 12645, there are periods of 5/23/37-hr over 18/6/5 cycles, respectively. AR 12470 also shows $\sim$7/9/19/39-hr periods which are observed through 23/18/9/4 cycles in the evolution of the EM flux, respectively. In the case of AR 11785, the SH and the total has a 23-hr period with 3 cycles only. 

 From Figures~\ref{Flare}--\ref{No-Flare} we conclude that shorter periods (5-10-hr) mostly appear when an AR is in the fully developed phase (e.g. ARs 12192 and 12470). At this stage, small amounts of flux appear or disappear but the total flux of an AR does not change dramatically. AR 11890 has shorter periods until the magnetic fluxes start to break apart and slowly dissipate. In the case of AR 12645, the 5-hr period becomes significant as the AR reached its fully developed phase. It seems that long-term periodicities are present during the entire lifetime of an AR.  

The WPS and GPS of the non-flaring ARs reveal no evidence for the 24- or 12-hr oscillations that are claimed to be present in HMI data due to the orbital motion of the SDO spacecraft. This effect has been reported by \cite{Liu2012}, where the Zeeman splitting coupled with the Doppler effect due to the Sun's rotation and the spacecraft motion causes the spectral line to shift every 12 and 24 hours. \cite{Smirnova2013a, Smirnova2013} found that the amplitude of these oscillations increases rapidly when the field strength exceeds 2000 G in the magnetic fields of ARs. \cite{Kutsenko2016} further argued for the presence of these two artificial oscillations by studying wavelet transform of the solar mean magnetic field measurements. If these oscillations were indeed significant in the helicity components presented in this work, we would expect to see them in the WPS of all ARs, particularly in the non-flaring regions where there are very few periodicities of significant power. If they are present, they are weak, and below the threshold significant levels.

\section{Summary and discussion} \label{conclusion}

Comparing the evolution of the helicity fluxes between the flaring and non-flaring ARs is important, since it reflects the dynamic evolution of an AR. The magnetic helicity is uniquely related to the geometrical complexity of the underlying magnetic system, determined by the twist and writhe of individual magnetic field lines, as well as their mutual entanglement. Therefore, the helicity plays an important role in solar activity phenomena, and, the magnetic-helicity-based quantities may be efficient for the purpose of flare prediction \citep[see e.g.][and references therein]{Pariat2017,Thalmann2019}. It remains still a challenging task to find an improved characterisation of the evolution of helicity injection inside an AR, and employ this information as a practical tool in the context of flare prediction. 

In this work, we determine the emergence, shearing, and total helicity components of six ARs by using the DAVE4VM algorithm \citep{Schuck2008}. Three ARs produced intensive solar flare eruptions and another three ARs were host of smaller B- and C-class flares only. In the case of flaring/non-flaring groups, one AR was selected to represent each of the three morphology phases of formation, fully developed, and decay. Following a wavelet analysis of the time series of normalised helicity flux components, we found the following:
 \begin{enumerate}
 \item Flaring ARs show common and rather powerful periodicities in the time series of the normalised emergence, shearing and total helicity fluxes. These common periodicities tend to  appear before the occurrence of the large flares. 
 \item Non-flaring ARs do not possess such clear common periodicities present in the three magnetic helicity components. 
 \item  Shorter periods, e.g. between 5-10-hr, are observable when an AR is in its fully developed evolutionary phase.
 \item Longer periods are present during an AR's lifetime. The identified longer periods are found to be comparable with the results of \cite{Goldvarg2005}. They found a 48-hr periodicity of the energy release of ARs in a larger statistical example. 
\end{enumerate}

The periodicity of EM and SH components of magnetic helicity may reflect the evolution of ARs where the magnetic flux emergence, the complexity evolution, and the subsequent energy release, do not occur monotonically but by alternating and periodic phases. Supporting \cite{Pariat2017} and \cite{Thalmann2019}, we can also conclude that the three helicity flux components are together capable to reveal the threat of a flaring AR, but not the magnitude of an upcoming eruption. Our findings demand a similar analysis on a much larger dataset to draw more firm conclusions about the flaring precursor capability and accuracy of helicity flux.

\section*{Acknowledgements} 
The authors are grateful to an anonymous Referee for constructive comments and recommendations which helped to improve the readability and quality of the paper. MBK and HM are grateful to the Science and Technology Facilities Council (STFC), (UK, Aberystwyth University, grant number ST/S000518/1), for the support received while conducting this research. All authors acknowledge the SOLARNET Mobility Programme  {\it grant number 824135}. RE is grateful to STFC (UK, grant number ST/M000826/1) and EU H2020 (SOLARNET, grant number 158538). RE also acknowledges support from the Chinese Academy of Sciences President’s International Fellowship Initiative (PIFI, grant number 2019VMA0052) and The Royal Society (grant nr IE161153). PR acknowledges support by the Italian MIUR-PRIN grant 2017APKP7T on \textit{Circumterrestrial Environment: Impact of Sun-Earth Interaction}.The research leading to these results has received funding from the European Union’s Horizon 2020 research and innovation programme under grant agreement no. 739500 (PRE-EST project) and no. 824135 (SOLARNET project). This work was supported by the Italian MIUR-PRIN 2017 on Space Weather: impact on circumterrestrial environment of solar activity, by Space Weather Italian COmmunity (SWICO) Research Program, and by the Università degli Studi di Catania (Piano per la Ricerca Università di Catania 2016-2018 – Linea di intervento 1 “Chance”; Linea di intervento 2 “Ricerca di Ateneo - Piano per la Ricerca 2016/2018”; “Fondi di ateneo 2020-2022, Università di Catania, linea Open Access”)
 
\FloatBarrier
\bibliography{adssample}

\newpage

 \begin{table*}
 \centering
\begin{tabular}{|c|cccc|}
\hline

AR &	Flare Class 	&	Flare Time	& Evolutionary phase& $\delta$-spot \\\hline
	\multicolumn{5}{|c|}{{\bf Flaring ARs}}		\\\hline

 {\bf AR 11166} 	&	X1.5	&	09/03/2011 23:23 &Emergence	&2011.03.05-03.11	\\
	2011.03.04-03.11&		&	&&	\\
			&		&	&	&	\\
{\bf AR 12192} 	&	X1.1	&	19/10/2014 05:03	&Stable&	2014.10.19 -10.27\\
2014.10.19 -10.27	&	M8.7	&	22/10/2014 01:59 &&		\\
	&	X1.6	&	22/10/2014  14:28	&&	\\
	&	X3.1	&	24/10/2014 21:41&&		\\
	&	X1.0	&	25/10/2014 17:08 &&  		\\
	&	X2.0	&	26/10/2014 10:56&&	\\
		&		&		&&		\\	
 {\bf AR 11890} 	&	X3.3	&	05/11/2013 22:12	&Decay&2013.11.04-11.12	\\
	2013.11.04-11.12&		X1.1	&	08/11/2013 04:26	&&	\\
	&	X1.1	&	10/11/2013 05:14	&&	\\ 
	\hline
	\multicolumn{5}{|c|}{{\bf Non-flaring ARs	}}		\\\hline
  {\bf AR 12645 }	&	B /C	&&Emergence	&	2017.04.02-04.06\\
 2017.03.29-04.07	&		&	&&	\\	
&		&		&&	\\
{\bf AR 12470 }	&	B /C	& &Stable	&	2015.12.15-12.16 	\\
 2015.12.15-12.22	&		&	&&		\\
 &		&		&&		\\	
{\bf  AR 11785} 	&	B/ C	&	&Decay&	2013.07.04-07.08\\
 2013.07.04-07.12 	&		&	&&		\\\hline
 \end{tabular}
 \caption{\label{ARsmagnetogramtable}  
Summary table of the properties of the studied six ARs: NOAA number and the study period of AR, information about the investigated flares, dominant evolution phase of the ARs. The last coulomb shows the time interval when $\delta$-spot(s) appeared in an AR.   }
\end{table*}

\end{document}